\documentclass{emulateapj}
\usepackage{amsmath}

\shorttitle{GRB Afterglow Spectrum}
\shortauthors{Fukushima et al.}

\begin{document}

\title{
Temporal Evolution of the Gamma-Ray Burst Afterglow Spectrum
for an Observer: GeV--TeV Synchrotron Self-Compton Light Curve
}

\author{Takuma Fukushima\altaffilmark{1},
Sho To\altaffilmark{2}, Katsuaki Asano\altaffilmark{2},
and Yutaka Fujita\altaffilmark{1}}

\affil{\altaffilmark{1}Department of Earth and Space Science,
Osaka University, Osaka, 560-0043, Japan; fukushima@vega.ess.sci.osaka-u.ac.jp,
fujita@vega.ess.sci.osaka-u.ac.jp}
\affil{\altaffilmark{2}Institute for Cosmic Ray Research, The University of Tokyo,
5-1-5 Kashiwanoha, Kashiwa, Chiba 277-8582, Japan;
tosho@icrr.u-tokyo.ac.jp, asanok@icrr.u-tokyo.ac.jp}


\begin{abstract}
We numerically simulate the gamma-ray burst (GRB) afterglow emission
with a one-zone time-dependent code.
The temporal evolutions of the decelerating shocked shell
and energy distributions
of electrons and photons are consistently calculated.
The photon spectrum and light curves for an observer
are obtained taking into account
the relativistic propagation of the shocked shell and the curvature
of the emission surface.
We find that the onset time of the afterglow
is significantly earlier than the previous analytical
estimate.
The analytical formulae of the shock propagation
and light curve
for the radiative case
are also different from our results.
Our results show that even if the emission mechanism
is switching from synchrotron to synchrotron self-Compton,
the gamma-ray light curves can be a smooth power-law,
which agrees with the observed light curve and
the late detection of a 32 GeV photon in GRB 130427A.
The uncertainty of the model parameters
obtained with the analytical formula is discussed,
especially in connection with the closure relation
between spectral index and decay index.
\end{abstract}


\section{Introduction}
\label{sec:intro}

The afterglow emission of gamma-ray bursts (GRBs)
is robust evidence of electron acceleration at
relativistic shocks.
While the difficulty of the particle acceleration
by magnetized relativistic shocks has been pointed out by
several authors \citep[e.g.][]{sir09,lem10,sir13}
from the theoretical point of view,
the low magnetization implied from recent broadband
observations of the afterglows \citep[e.g.][]{kum09,lem13,san14,ben15,zha15}
seems to be consistent with the theoretical argument.

The physical property of the shock and electron acceleration
in the GRB afterglows has been investigated with the conventional microscopic
parameters,
the energy fractions of the accelerated electrons $\epsilon_{\rm e}$
and magnetic field $\epsilon_B$ in the downstream.
The observations and standard analytical formulae
of the external shock model
by \citet{sar98} provide those microscopic parameters
and jet parameters \citep[e.g.][]{pan02,llo04}.
In those analytical formulae, the electron energy distribution
at a given radius is assumed to be a broken-power-law.
The results of the two-dimensional hydrodynamical simulations by \citet{vEer12}
with the analytical broken-power-law formula
have been widely used to fit the observed light curves \citep[see, e.g.][]{gui14,mas14,zha15}.
Such multidimensional hydrodynamical simulations provide
precise evolution of the shock propagation and angular structure of the collimated jet
and are a powerful tool to constrain jet parameters,
especially in the off-axis cases.

On the other hand, the actual electron energy distribution
in the downstream of the propagating shock is not a simple broken-power-law.
\citet{pet09} and \citet{pen14} calculated the evolution of the electron energy distribution
in the afterglow.
The resultant photon spectra are significantly curved around the
cooling and injection break frequencies, and not the broken-power-law
\citep[see also][]{uhm14}.

Some fraction of GRB afterglows are hard to explain
with the standard external shock model
\citep[e.g.][]{wil07,wan15}.
Multizone models such as the spine-sheath structure \citep{rac08},
the contribution of the reverse shock \citep{gen07,uhm07},
or evolving microscopic parameters \citep{iok06} may be required
to reconcile such exceptional afterglows.
Before increasing the number of parameters following such complex models,
however, we need to clarify the degree of the contradiction with the standard
external shock model.
In addition to the uncertainty of the electron energy distribution,
the detections of the GeV afterglows with {\it Fermi} \citep{abd09,kum10}
require us to investigate seriously the effect of synchrotron self-Compton
(SSC) emission on the spectrum and light curve.
Especially, the detection of a 32 GeV photon
at $3 \times 10^4$ s in GRB 130427A \citep{ack14}
cannot be explained with the usual synchrotron emission
for the standard evolution of the external shock.
The SSC emission spectra numerically obtained \citep{pet09,pen14}
are naturally different from a broken-power-law
derived from the analytical description \citep{sar01}.
In addition, if the shocked plasma is in the highly radiative regime
as discussed in \citet{ghi10}, the radiative cooling affects not only
the electron energy distribution but also the evolution of the bulk Lorentz factor.
When we treat all of the above nontrivial effects numerically
without analytical approximations,
the spectrum and light curve may deviate from the behaviors
given by simple formulae.

In this paper, in order to discuss the uncertainty of the evolution of the emission
from the external shock,
we simulate the evolutions of the shocked material propagating in the interstellar medium (ISM).
Our numerical code is based on the one-zone approximation, but the time-dependent treatment
is completely applied for the bulk motion of the shell
and the electron and photon energy distributions.
Our method is similar to that in the previous studies \citep{pet09,pen14,uhm14},
but the light curves were not calculated in their studies.
Our code consistently transforms the energy and arrival time of photons that
escaped from the shocked shell into those for an observer.
The spectrum for the observer at a certain time $t_{\rm obs}$
is not just the blue-shifted one in the shell comoving frame
at the time $t'$ given by the one-to-one correspondence
between $t_{\rm obs}$ and $t'$.
Focusing on the light curve and spectral evolution for the observer,
we discuss the differences in the results obtained
with the analytical method and ours.
We also discuss the switching signature from synchrotron
to inverse Compton in the gamma-ray light curve.
In \S \ref{sec:form}, we present our
computing method.
The analytical formulae in \S \ref{sec:ana} are compared with
the numerical results in \S \ref{sec:res}.
Our model is applied to the afterglow of GRB 130427A in \S \ref{sec:130}.
The smooth gamma-ray light curve is reproduced in spite of
the switching of the emission process in the GeV energy range.
The conclusions are summarized in \S \ref{sec:conc}.

\section{Model and Method}
\label{sec:form}

In this paper, we assume a spherically symmetric system,
which may be an appropriate assumption before the jet break.
We treat the shocked region propagating in the ISM
as a uniform shell with a thickness $W=W'/\Gamma$,
where $\Gamma \equiv 1/\sqrt{1-\beta^2}$ is the bulk Lorentz factor of the shocked region.
Hereafter, we denote values in the shell frame by primed characters.
Under the one-zone approximation, we numerically solve
the evolutions of $\Gamma$,
magnetic field, and energy distributions of the photons
and non-thermal electrons in the shell in a self-consistent manner.
The model parameters are the total energy $E_0$ promptly released
from the central engine, the initial bulk Lorentz factor of the ejecta
$\Gamma_0$, the proton number density of the ISM $n_{\rm ISM}$,
the spectral index $p$ and 
the number fraction $\eta$ of non-thermal electrons,
and the energy fractions 
to the shock-dissipated energy, $\epsilon_{\rm e}$ and $\epsilon_B$,
of non-thermal electrons and magnetic field, respectively.
Below, we present the evolution of the shell regardless of whether
the shell motion is relativistic or not.
The average kinetic energy per proton just behind the shock front
is $(\Gamma-1) m_{\rm p} c^2$; hence, we can obtain the temperature
$T_{\rm sh}$ from
\begin{equation}
\Gamma=\frac{K_3(1/x)}{K_2(1/x)}-x,
\label{temp}
\end{equation}
where $x \equiv T_{\rm sh}/(m_{\rm p} c^2)$
and $K_n(x)$ is the modified Bessel function of the second kind.
Given $x$ and $\Gamma$, the heat capacity ratio is written as
\begin{equation}
\hat{\gamma}=\frac{x}{\Gamma-1}+1.
\label{hcap}
\end{equation}
The shock jump condition \citep{bla76} provides the bulk Lorentz factor of the shock front
$\Gamma_{\rm sh}$ as
\begin{equation}
\Gamma_{\rm sh}=\sqrt{\frac{(\Gamma+1)(\hat{\gamma}(\Gamma-1)+1)^2}
{\hat{\gamma} (2-\hat{\gamma})(\Gamma-1)+2}}.
\label{gsh}
\end{equation}

When the shock front is propagating at a radius $R$ from the central engine as
$dR/dt=c \beta_{\rm sh}$ ($dt=\Gamma dt'$, $\beta_{\rm sh} \equiv
\sqrt{1-1/\Gamma_{\rm sh}^2}$),
the mass in the shell evolves as
\begin{equation}
\frac{dM}{dt}=\frac{1}{\Gamma}\frac{dM}{dt'}=4 \pi R^2 c \beta_{\rm sh} n_{\rm ISM} m_{\rm p},
\label{mdot}
\end{equation}
with the initial mass $M_0=E_0/(\Gamma_0-1)/c^2$.
The total energy including the rest mass energy in the comoving frame
evolves as
\begin{equation}
\frac{dE'_{\rm sh}}{dt'}=\Gamma c^2
\frac{dM}{dt'}-\frac{dE'_{\rm rad}}{dt'}-\frac{dE'_{\rm ad}}{dt'},
\label{edot}
\end{equation}
where the first through third terms on the right-hand side express the energy injection,
radiative cooling, and adiabatic cooling, respectively.
For each time step,
we numerically follow the evolutions of the shell mass and energy with Equations
(\ref{mdot}) and (\ref{edot})
and obtain $\Gamma$ from the energy conservation
\begin{equation}
E_{\rm sh}=\Gamma E'_{\rm sh}=E_0 +M c^2-E_{\rm rad}.
\label{econ}
\end{equation}

Since we assume a homogeneous shell,
the density obtained by the jump condition,
\begin{equation}
n'=\frac{\hat{\gamma} \Gamma+1}{\hat{\gamma}-1}n_{\rm ISM},
\label{nden}
\end{equation}
is adopted for the entire shell.
According to the evolutions of $M$ and $\Gamma$,
the shell volume is written as $V'=M/(m_{\rm p} n')$.
Assuming that a fraction $\epsilon_B$ of the injected kinetic energy
converts into the magnetic field,
the magnetic energy $E_B$ is calculated by
\begin{equation}
\frac{dE'_B}{dt'}=\epsilon_B (\Gamma-1) c^2
\frac{dM}{dt'}.
\label{Bdot}
\end{equation}
The magnetic field is estimated by
\begin{equation}
B'=\sqrt{\frac{8\pi E'_B}{V'}}.
\label{mag}
\end{equation}

The evolution of the electron and photon energy distributions
in the shell frame is calculated with the same method as
in \citet{asa11}.
Non-thermal electrons (number fraction $\eta$) are assumed to obtain
a fraction $\epsilon_{\rm e}$ of the injected kinetic energy.
Assuming a cut-off power-law spectrum at injection
\begin{equation}
\dot{N'}_{\rm inj}(\varepsilon'_{\rm e}) =\dot{N'_0}
(\varepsilon'_{\rm e}/\varepsilon'_{\rm min})^{-p}
\exp{(-\varepsilon'_{\rm e}/\varepsilon'_{\rm max})},
\label{elef}
\end{equation}
for $\varepsilon' \geq \varepsilon'_{\rm min}$,
the number and energy injection rates are written as
\begin{eqnarray}
\frac{dN'_{\rm e}}{dt'}&=&\frac{\eta}{m_{\rm p}}
\frac{dM}{dt'}
=\int_{\varepsilon'_{\rm min}}^\infty d \varepsilon'
\dot{N'_{\rm e}}(\varepsilon'_{\rm e}) \label{eNdot}, \\
\frac{dE'_{\rm e}}{dt'}&=&\epsilon_{\rm e} (\Gamma-1) c^2
\frac{dM}{dt'}
=\int_{\varepsilon'_{\rm min}}^\infty d \varepsilon'
\varepsilon_{\rm e}' \dot{N'_{\rm e}}(\varepsilon'_{\rm e}). \label{eEdot}
\end{eqnarray}
The maximum electron energy $\varepsilon'_{\rm max}$
is obtained by equating the acceleration time $\xi r'_{\rm L}/c$
(or $20 \xi r'_{\rm L}/(3 \beta^2 c)$ for the non-relativistic case) and
cooling time $\varepsilon_{\rm e}/\dot{\varepsilon_{\rm e}}$
due to synchrotron and inverse Compton emissions
numerically obtained, where $r_{\rm L}$ is the Larmor radius.
Hereafter, the Bohm factor $\xi$ is optimistically assumed as unity.
Then, Equations (\ref{eNdot}) and (\ref{eEdot}) provide
the normalization $\dot{N'_0}$ and $\varepsilon'_{\rm min}$
for given $dM/dt'$, $\Gamma$, $\eta$, and $\epsilon_{\rm e}$.

In this paragraph, to explain the method for following the
evolution of the electron/positron/photon energy
distributions, we omit the prime symbol and express equations
in the shell frame. Our numerical code practically solves
the evolution equation of non-thermal electrons/positrons
\begin{eqnarray}
\frac{\partial N_{\rm e}(\varepsilon_{\rm e})}{\partial t}=
\frac{\partial }{\partial \varepsilon_{\rm e}}
\Bigl[ \Bigl( \left< \dot{\varepsilon_{\rm e}} \right>_{\rm syn}
+\left< \dot{\varepsilon_{\rm e}} \right>_{\rm IC}
+\left< \dot{\varepsilon_{\rm e}} \right>_{\rm ad}
\nonumber \\
-\left< \dot{\varepsilon_{\rm e}} \right>_{\rm SSA} \Bigr)
N_{\rm e}(\varepsilon_{\rm e})
\Bigr] +\dot{N}_{{\rm e},\gamma \gamma}(\varepsilon_{\rm e})
+\dot{N}_{\rm inj}(\varepsilon_{\rm e}),
\end{eqnarray}
where $\left< \dot{\varepsilon_{\rm e}} \right>_{\rm syn}$
and $\left< \dot{\varepsilon_{\rm e}} \right>_{\rm IC}$
are the energy loss rates (positive values)
due to synchrotron and inverse Compton (IC) emissions,
respectively.
The Klein--Nishina effect is numerically taken into account
using the table of the emissivity prepared in advance
with the Monte Carlo method \citep[see][]{asa11}.
The electron heating rate due to the synchrotron self-absorption (SSA)
is also included as denoted with $\left< \dot{\varepsilon_{\rm e}} \right>_{\rm SSA}$.
The extra term of electron--positron pair injection due to $\gamma \gamma$-absorption
is $\dot{N}_{{\rm e},\gamma \gamma}(\varepsilon_{\rm e})$.
The adiabatic cooling term $\left< \dot{\varepsilon_{\rm e}} \right>_{\rm ad}$
is calculated from the momentum evolution
$\dot{p_{\rm e}}=-p_{\rm e} \dot{V}/(3V)$.
Since the kinetic energy is
$\varepsilon_{\rm e}=\sqrt{p_{\rm e}^2 c^2+m^2_{\rm e} c^4}-m_{\rm e} c^2$,
the cooling rate is written as
\begin{eqnarray}
\left< \dot{\varepsilon_{\rm e}} \right>_{\rm ad}
=\frac{1}{3}\frac{\dot{V}}{V}\frac{\varepsilon_{\rm e}^2+2 \varepsilon_{\rm e}m_{\rm e} c^2}
{\varepsilon_{\rm e}+m_{\rm e} c^2}.
\label{adc}
\end{eqnarray}
The pair production, IC emission, and SSA depend on
the photon density $n_\gamma(\varepsilon)=N_\gamma(\varepsilon)/V$.
The photon energy distribution is obtained by solving
\begin{eqnarray}
\frac{\partial N_{\gamma}(\varepsilon)}{\partial t}=
\dot{N}_{\gamma,{\rm syn}}(\varepsilon)+\dot{N}_{\gamma,{\rm IC}}(\varepsilon)
-\dot{N}_{\gamma,{\gamma \gamma}}(\varepsilon) \nonumber \\
-\dot{N}_{\gamma,{\rm SSA}}(\varepsilon)-\dot{N}_{\gamma,{\rm esc}}(\varepsilon),
\end{eqnarray}
where the first and second terms on the right-hand side represent
synchrotron and IC photon production, respectively,
and the third and fourth terms represent photon absorption due to $\gamma \gamma$
and SSA, respectively.
Those terms are numerically calculated with the given electron and photon
distributions and magnetic field.
Photons escape from both the front and rear surfaces, so that
the escape term is written as
\begin{eqnarray}
\dot{N}_{\gamma,{\rm esc}}(\varepsilon)=\frac{c}{2W} N_{\gamma}(\varepsilon),
\end{eqnarray}
where the shell width $W=V/(4 \pi R^2)$.

Using the prime symbol again hereafter,
the radiative cooling term in Equation (\ref{edot}) is obtained as
\begin{eqnarray}
\frac{dE'_{\rm rad}}{dt'}=
\int d\varepsilon' \dot{N'}_{\gamma,{\rm esc}}(\varepsilon')\varepsilon',
\label{radev}
\end{eqnarray}
and the radiation term in Equation (\ref{econ}) is calculated with
\begin{eqnarray}
E_{\rm rad}=\int dt' \Gamma \frac{dE'_{\rm rad}}{dt'}.
\end{eqnarray}
Although we do not solve the proton energy distribution explicitly,
the adiabatic cooling of protons is essential for the evolution of $\Gamma$.
The energy injection rate into protons is $dE'_{\rm p}/dt'=(1-\epsilon_{\rm e}-\epsilon_B)
(\Gamma-1)c^2 dM/dt'$.
The average kinetic energy of protons $\bar{\varepsilon}'_{\rm p}
\equiv m_{\rm p} E'_{\rm p}/M$ evolves as
\begin{eqnarray}
\frac{d\bar{\varepsilon}'_{\rm p}}{dt'}
&=&\left\{ (1-\epsilon_{\rm e}-\epsilon_B) (\Gamma-1)c^2 -\frac{E'_{\rm p}}{M} \right\}
\frac{m_{\rm p}}{M} \frac{dM}{dt'} \nonumber \\
&&-\left< \dot{\bar{\varepsilon}}_{\rm p} \right>_{\rm ad},
\end{eqnarray}
where the last term is the same form as Equation (\ref{adc}) with ${\rm e} \to {\rm p}$.
This simplified method provides the adiabatic energy loss rate
\begin{eqnarray}
\frac{dE'_{\rm ad}}{dt'}=\frac{M}{m_{\rm p}}
\left< \dot{\bar{\varepsilon}}_{\rm p} \right>_{\rm ad}
+\int d\varepsilon'_{\rm e} N'_{\rm e}(\varepsilon'_{\rm e})
\left< \dot{\varepsilon_{\rm e}} \right>_{\rm ad}.
\label{adev}
\end{eqnarray}
With Equations (\ref{mdot}), (\ref{radev}), and (\ref{adev}),
the total shell energy is calculated from Equation (\ref{edot}).
Then, we can obtain the Lorentz factor $\Gamma$ from Equation
(\ref{econ}) for each time step.

In order to obtain the photon spectrum and light curve
for an observer, we integrate photons over the shell surface.
The method for the time and energy transformations is also the same
as in \citet{asa11}.
The energy and arrival time of photons escaping from 
the surface expanding toward an angle $\theta$
to the line of sight at radius $R$
are written as
\begin{eqnarray}
\varepsilon_{\rm obs}&=&\frac{\varepsilon'}{\Gamma(1-\beta \cos{\theta})(1+z)},
\label{etrf} \\
t_{\rm obs}&=&(1+z) \Bigl[ (t-\frac{R-R_0}{c}\cos{\theta})
+\frac{R_0}{c}(1-\cos{\theta}) \Bigr], \nonumber \\
\label{ttrf}
\end{eqnarray}
where $t=\int \Gamma dt'$, $R=c \int \beta_{\rm sh} \Gamma dt'$,
and $R_0$ is the initial radius.
In the comoving frame, the photon escape rate per unit surface
per solid angle is written as
\begin{eqnarray}
\frac{dN'_\gamma}{d \Omega' dS' dt' d \varepsilon'}=c|\cos{\theta'}|
\frac{n'_\gamma(\varepsilon',t')}{4 \pi},
\end{eqnarray}
where $dS'=dS=2 \pi R^2 \sin{\theta} d \theta$ is the surface element.
While the number of photons $dN'_\gamma$ is obviously Lorentz invariant,
the infinitesimal intervals are transformed as
$dt_{\rm obs} =(1+z)(1-\beta_{\rm sh} \cos{\theta})\Gamma dt'$,
$d \varepsilon_{\rm obs}=d \varepsilon'/\left\{(1+z)(1-\beta \cos{\theta})\Gamma\right\}$,
and $d \Omega=\Gamma^2 (1-\beta \cos{\theta})^2 d \Omega'$ for solid angle.
Denoting the luminosity distance as $D_{\rm L}=(1+z)D$,
the surface through which photons traveling toward $d \Omega$ pass
is written as $dS_{\rm obs}=D^2 d \Omega$. Then,
we obtain the photon flux for an observer
as
\begin{eqnarray}
&&\Phi(\varepsilon_{\rm obs},t_{\rm obs})
= \frac{dN_\gamma}{dS_{\rm obs} dt_{\rm obs} d \varepsilon_{\rm obs}}=\nonumber \\
&& \int d\theta \left( \frac{R(t'_\theta)}{D} \right)^2
\frac{\sin{\theta}|\cos{\theta'}| c n'_\gamma(\varepsilon'_\theta,t'_\theta)}
{2\Gamma^2 (1-\beta \cos{\theta})(1-\beta_{\rm sh} \cos{\theta})},
\label{obsf}
\end{eqnarray}
where
\begin{eqnarray}
\cos{\theta'}=\frac{\cos{\theta}-\beta}{1-\beta \cos{\theta}},
\end{eqnarray}
and $\varepsilon'_\theta$ and $t'_\theta$ are the comoving energy and time obtained
from Equations (\ref{etrf}) and (\ref{ttrf}), respectively,
for given $\theta$, $\varepsilon_{\rm obs}$, and $t_{\rm obs}$.
Notice that $\beta$ (equivalently $\Gamma$)
and $\beta_{\rm sh}$ are also functions of $t'_\theta$.
Carrying out the integral in Equation (\ref{obsf}) numerically over $\theta$,
we can obtain the spectral evolution
for an observer.

\section{Analytical Behavior: Review}
\label{sec:ana}

While we numerically follow the evolution of the photon spectrum
for an observer with the method
explained in the previous section,
here we review the analytical description in \citet{sar98}
to compare with our results.
When the shock is ultra-relativistic ($\Gamma \gg 1$),
$\hat{\gamma} \simeq 4/3$ and $\Gamma_{\rm sh} \simeq \sqrt{2} \Gamma$.
Until the deceleration radius \citep{ree92},
\begin{eqnarray}
R_{\rm dec}\simeq\left( \frac{3E_0}{4 \pi n_{\rm ISM} m_{\rm p} c^2 \Gamma_0^2}
\right)^{1/3},
\label{rdec}
\end{eqnarray}
the shell expands with a constant Lorentz factor $\Gamma_0$.
The peak time of the afterglow for an observer corresponds to this radius as
\begin{eqnarray}
t_{\rm obs,pk}\simeq(1+z)\frac{R_{\rm dec}}{2c \Gamma_0^2}
\simeq 90 (1+z)E_{52}^{\frac{1}{3}} n_0^{-\frac{1}{3}} \Gamma_2^{-\frac{8}{3}}
~\mbox{s},
\label{tpk}
\end{eqnarray}
where $E_0=10^{52} E_{52}~\mbox{erg}$, $n_{\rm ISM}=n_0~\mbox{cm}^{-3}$,
and $\Gamma_0=100 \Gamma_2$.
When the peak time is determined observationally,
the initial Lorentz factor is estimated as
\begin{eqnarray}
\Gamma_0 \simeq 96 E_{52}^{\frac{1}{8}} n_0^{-\frac{1}{8}} t_{\rm pk,2}^{-\frac{3}{8}},
\label{g0}
\end{eqnarray}
where $t_{\rm obs,pk}/(1+z)=100 t_{\rm pk,2}$ s.
After the peak time, the shell starts to decelerate.
Since the shell density is $n'\simeq 4 \Gamma n_{\rm ISM}$,
the shell width becomes $W \simeq R/(12 \Gamma^2)$ in the one-zone approximation.
The jump condition provides the energy density $U'=4 \Gamma^2 n_{\rm ISM} m_{\rm p} c^2$.
Neglecting the radiative cooling, the energy conservation
implies that the Lorentz factor decreases as
\begin{eqnarray}
\Gamma\simeq\sqrt{\frac{3E_0}{4 \pi n_{\rm ISM} m_{\rm p}c^2R^3}}.
\end{eqnarray}
The one-zone approximation in the above equation
has a slightly different factor from that in \citet{sar98},
where the radial density structure behind the shock is taken into account
to estimate $E_0$.
The simple one-to-one correspondence for $R$ and $t_{\rm obs}$,
$t_{\rm obs} \simeq (1+z) R/(4 c \Gamma^2)$, implies
\begin{eqnarray}
R \simeq 1.6 \times 10^{17} (1+z)^{-\frac{1}{4}} n_0^{-\frac{1}{4}}
E_{52}^{\frac{1}{4}}
t_{\rm h}^{\frac{1}{4}}~\mbox{cm},
\end{eqnarray}
where $t_{\rm obs}=t_{\rm h}$ hr.

From Equations (\ref{eNdot}) and (\ref{eEdot}), we obtain
the electron minimum Lorentz factor
$\gamma'_{\rm m}-1 \equiv \varepsilon'_{\rm min}/(m_{\rm e} c^2)$ as
\begin{eqnarray}
\gamma'_{\rm m} \simeq \frac{\epsilon_{\rm e}}{\eta}
\frac{p-2}{p-1} \Gamma \frac{m_{\rm p}}{m_{\rm e}}.
\end{eqnarray}
A fraction $\epsilon_B$ of the energy density converts to the magnetic field as
\begin{eqnarray}
B'\simeq\Gamma \sqrt{32 \pi \epsilon_B n_{\rm ISM} m_{\rm p} c^2}.
\label{B-field}
\end{eqnarray}
The typical synchrotron photon energy is obtained as
\begin{eqnarray}
\varepsilon_{\rm m}&\simeq& \frac{\Gamma}{1+z}
\frac{3 \hbar e B'}{2 m_{\rm e} c} \gamma'^2_{\rm m} \\
&\simeq& 0.28 (1+z)^{\frac{1}{2}} f_{1/6}^2 \eta^{-2} \epsilon_{{\rm e},-1}^2
\epsilon_{B,-1}^{\frac{1}{2}} E_{52}^{\frac{1}{2}}
t_{\rm h}^{-\frac{3}{2}}~\mbox{eV},
\label{em}
\end{eqnarray}
where $(p-2)/(p-1)=1/6 f_{1/6}$,
$\epsilon_{\rm e}=0.1\epsilon_{{\rm e},-1}$, and
$\epsilon_B=0.1\epsilon_{B,-1}$.
Given the photon energy $\varepsilon_{\rm obs}=\varepsilon_{\rm eV}$ eV
in observation,
$\varepsilon_{\rm m}$ passes at an observer time
\begin{eqnarray}
t_{\rm obs,m}&\simeq& 
1500 (1+z)^{\frac{1}{3}} f_{1/6}^{\frac{4}{3}} \eta^{-\frac{4}{3}}
\epsilon_{{\rm e},-1}^{\frac{4}{3}}
\epsilon_{B,-1}^{\frac{1}{3}} E_{52}^{\frac{1}{3}}
\varepsilon_{\rm eV}^{-\frac{2}{3}}~\mbox{s}. \nonumber \\
\end{eqnarray}

In the electron energy distribution,
the cooling break appears at a Lorentz factor
\begin{eqnarray}
\gamma'_{\rm c} \simeq \frac{6\pi (1+z)m_{\rm e} c}
{\sigma_{\rm T} B'^2 \Gamma t_{\rm obs}}.
\label{eqgc}
\end{eqnarray}
Note that the formulation in this section neglects the effect
of IC cooling.
When IC is dominant for electron cooling,
$\gamma'_{\rm c}$ and its evolution will be modified.
In our numerical simulations shown in the next section, those non-linear effects due to IC
are automatically included. From Equation (\ref{eqgc}),
the cooling break energy becomes
\begin{eqnarray}
\varepsilon_{\rm c}&\simeq& 
3.1 (1+z)^{-\frac{1}{2}}
\epsilon_{B,-1}^{-\frac{3}{2}} E_{52}^{-\frac{1}{2}} n_0^{-1}
t_{\rm h}^{-\frac{1}{2}}~\mbox{eV}.
\label{ec}
\end{eqnarray}
The radius corresponding to $\varepsilon_{\rm c}=\varepsilon_{\rm obs}$ is written as
\begin{eqnarray}
R_{\rm c}&\simeq& 
2.8 \times 10^{17} (1+z)^{-\frac{1}{2}}
\epsilon_{B,-1}^{-\frac{3}{4}} n_0^{-\frac{3}{4}}
\varepsilon_{\rm eV}^{-\frac{1}{2}}~\mbox{cm},
\end{eqnarray}
which corresponds to the observer time
\begin{eqnarray}
t_{\rm obs,c}&\simeq& 
3.2 \times 10^{4} (1+z)^{-1}
\epsilon_{B,-1}^{-3} n_0^{-2} E_{52}^{-1}
\varepsilon_{\rm eV}^{-2}~\mbox{s}.
\end{eqnarray}
According to the high/low relation between $\varepsilon_{\rm m}$ and $\varepsilon_{\rm c}$,
the non-thermal electrons are judged as the fast cooling
($\varepsilon_{\rm m}>\varepsilon_{\rm c}$) or the slow cooling
($\varepsilon_{\rm m}<\varepsilon_{\rm c}$).
In the early stage, the strong synchrotron cooling may lead to the fast cooling.
The transition from the fast cooling to the slow cooling
occurs at
\begin{eqnarray}
t_{\rm eq} \equiv R_{\rm eq}/c
\simeq 3.0 \times 10^6 f_{1/6}^{\frac{1}{2}}
\eta^{-\frac{1}{2}} \epsilon_{{\rm e},-1}^{\frac{1}{2}}
\epsilon_{B,-1}^{\frac{1}{2}} E_{52}^{\frac{1}{2}}
~\mbox{s},
\label{teq}
\end{eqnarray}
which corresponds to
\begin{eqnarray}
t_{\rm obs,eq}\simeq 330 (1+z) f_{1/6}^2 \eta^{-2} \epsilon_{{\rm e},-1}^2
\epsilon_{B,-1}^{2} E_{52} n_0
~\mbox{s},
\label{teq-obs}
\end{eqnarray}
for the observer.
To obtain the SSA frequency,
assuming that all the injected electrons form a
single power-law above $\min(\gamma'_{\rm m},\gamma'_{\rm c})$,
we calculate the usual absorption formula with the synchrotron
function.
When $t_{\rm obs}<t_{\rm eq}$,
the SSA frequency is
\begin{eqnarray}
\nu_{\rm a}&\simeq& 20 (1+z)^{-\frac{1}{2}}\eta^{\frac{3}{5}}
\epsilon_{B,-1}^{\frac{6}{5}} E_{52}^{\frac{7}{10}} n_0^{\frac{11}{10}}
t_{2}^{-\frac{1}{2}}~\mbox{GHz},
\end{eqnarray}
where $t_{\rm obs}=100 t_2$ s.
For $t_{\rm obs}>t_{\rm eq}$, we obtain a constant value
\begin{eqnarray}
\nu_{\rm a}&\simeq& 12 (1+z)^{-1} \frac{\Pi_p}{4.35}
\eta^{\frac{8}{5}} \epsilon_{{\rm e},-1}^{-1}
\epsilon_{B,-1}^{\frac{1}{5}} E_{52}^{\frac{1}{5}} n_0^{\frac{3}{5}}
~\mbox{GHz},
\end{eqnarray}
where
\begin{eqnarray}
\Pi_p \equiv \frac{p-1}{p-2}
\left( \frac{(p+2)(p-1)}{3p+2} \right)^{\frac{3}{5}},
\end{eqnarray}
which is $\sim 4.35$ for $p=2.2$.

As shown in \citet{sar98}, the maximum flux, $F_{\rm max}=\varepsilon_{\rm c}
\Phi(\varepsilon_{\rm c})$ for $t_{\rm obs}<t_{\rm eq}$ or
$\varepsilon_{\rm m} \Phi(\varepsilon_{\rm m})$ for $t_{\rm obs}>t_{\rm eq}$,
becomes constant for $t_{\rm obs}>t_{\rm obs,pk}$ as
\begin{eqnarray}
F_{\rm max}&\simeq& (1+z) \frac{N_{\rm e}}{4 \pi D_{\rm L}^2}
\frac{\sqrt{3} e^3 B'}{16 \hbar m_{\rm e} c^2} \Gamma \\
&\simeq& 1.4 \times 10^{-11} (1+z) \eta
\epsilon_{B,-1}^{\frac{1}{2}} E_{52} n_0^{\frac{1}{2}} D_{28}^{-2}
 \nonumber \\
&&\qquad \qquad \qquad \qquad \mbox{erg}~\mbox{cm}^{-2}~\mbox{s}^{-1}~\mbox{eV}^{-1},
\label{Fmax}
\end{eqnarray}
where $D_{\rm L}=10^{28} D_{28}$ cm.
Normalizing the flux $F(\varepsilon_{\rm obs})=\varepsilon_{\rm obs}
\Phi(\varepsilon_{\rm obs})$ by $F_{\rm max}$ at
$\min(\varepsilon_{\rm m},\varepsilon_{\rm c})$,
the broken-power-law formula yields
the spectral evolution as follows:
\begin{eqnarray}
F(\varepsilon_{\rm obs})\propto \begin{cases}
\varepsilon_{\rm obs}^2 t_{\rm obs} & \mbox{for}~\varepsilon_{\rm obs}<h \nu_{\rm a} \\
\varepsilon_{\rm obs}^{\frac{1}{3}} t_{\rm obs}^{\frac{1}{6}} &
\mbox{for}~h \nu_{\rm a} <\varepsilon_{\rm obs}<\varepsilon_{\rm c}\\
\varepsilon_{\rm obs}^{-\frac{1}{2}} t_{\rm obs}^{-\frac{1}{4}} &
\mbox{for}~\varepsilon_{\rm c} <\varepsilon_{\rm obs}<\varepsilon_{\rm m}\\
\varepsilon_{\rm obs}^{-\frac{p}{2}} t_{\rm obs}^{-\frac{3p-2}{4}} &
\mbox{for}~\varepsilon_{\rm m} <\varepsilon_{\rm obs},
\end{cases}
\label{eqsfast}
\end{eqnarray}
for $t_{\rm obs}<t_{\rm obs,eq}$, and
\begin{eqnarray}
F(\varepsilon_{\rm obs})\propto \begin{cases}
\varepsilon_{\rm obs}^2 t_{\rm obs}^{\frac{1}{2}} & \mbox{for}~\varepsilon_{\rm obs}<h \nu_{\rm a} \\
\varepsilon_{\rm obs}^{\frac{1}{3}} t_{\rm obs}^{\frac{1}{2}} &
\mbox{for}~h \nu_{\rm a} <\varepsilon_{\rm obs}<\varepsilon_{\rm m}\\
\varepsilon_{\rm obs}^{-\frac{p-1}{2}} t_{\rm obs}^{-\frac{3(p-1)}{4}} &
\mbox{for}~\varepsilon_{\rm m} <\varepsilon_{\rm obs}<\varepsilon_{\rm c}\\
\varepsilon_{\rm obs}^{-\frac{p}{2}} t_{\rm obs}^{-\frac{3p-2}{4}} &
\mbox{for}~\varepsilon_{\rm c} <\varepsilon_{\rm obs},
\end{cases}
\label{Fl-ana}
\end{eqnarray}
for $t_{\rm obs}>t_{\rm obs,eq}$.
In this section, we do not consider the cases of
$\min(\varepsilon_{\rm m},\varepsilon_{\rm c}) <h \nu_{\rm a}$,
in which case the spectral shape and its evolution should be modified
\citep[e.g.][]{gra02}.
For the parameter regions adopted in our simulations (see the next section),
the self-absorption frequency is safely suppressed below
$\min(\varepsilon_{\rm m},\varepsilon_{\rm c})$.

Before the peak time $t_{\rm pk}$,
$\Gamma$ and $B'$ are constant,
so that the maximum flux increases as $F_{\rm max} \propto N_{\rm e}
\propto R^3 \propto t_{\rm obs}^3$.
The characteristic photon energies behave as
$\varepsilon_{\rm m} \propto t_{\rm obs}^0$
and $\varepsilon_{\rm c} \propto t_{\rm obs}^{-2}$.
In the fast cooling case, $F(\varepsilon_{\rm obs})=F_{\rm max}
(\varepsilon_{\rm m}/\varepsilon_{\rm c})^{-1/2}
(\varepsilon_{\rm obs}/\varepsilon_{\rm m})^{-p/2}$ for
$\varepsilon_{\rm obs}>\varepsilon_{\rm m}$,
and $F(\varepsilon_{\rm obs})=F_{\rm max}
(\varepsilon_{\rm obs}/\varepsilon_{\rm c})^{-1/2}$ for
$\varepsilon_{\rm c}<\varepsilon_{\rm obs}<\varepsilon_{\rm m}$.
Then, as long as $\varepsilon_{\rm obs}>\varepsilon_{\rm c}$, we obtain
\begin{eqnarray}
F(\varepsilon_{\rm obs}) \propto F_{\rm max} \varepsilon_{\rm c}^{1/2}
\propto t_{\rm obs}^2,
\label{eqeLC}
\end{eqnarray}
for $t_{\rm obs}<t_{\rm obs,pk}$.
Similarly, for $\varepsilon_{\rm obs}<\varepsilon_{\rm c}$,
\begin{eqnarray}
F(\varepsilon_{\rm obs}) \propto F_{\rm max} \varepsilon_{\rm c}^{-1/3}
\propto t_{\rm obs}^{11/3}.
\label{eqeLC2}
\end{eqnarray}

\section{Numerical Results: Spectrum and Light Curve}
\label{sec:res}

\begin{figure}[!htb]
\centering
\epsscale{1.0}
\plotone{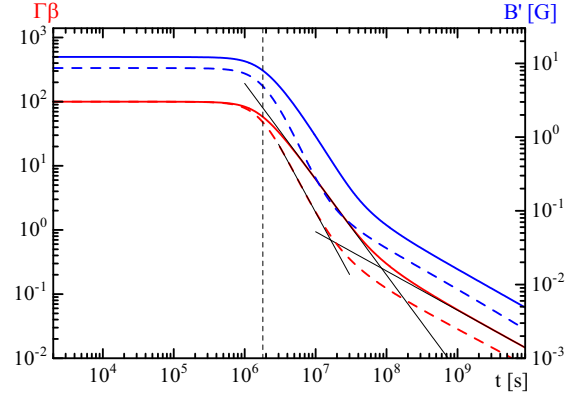}
\caption{Evolutions of $\Gamma \beta$ (red, left axis) and $B'$
(blue, right axis)
for the benchmark (solid) and radiative (dashed) parameter sets.
The vertical dashed line shows the deceleration time
$R_{\rm dec}/c$.
As the thin solid lines show, for the benchmark case,
the deceleration is consistent
with the power-law behavior $\Gamma \propto t^{-3/2}$
in the relativistic regime and $\beta \propto t^{-3/5}$
in the non-relativistic regime.
For the radiative case, the numerical result shows $\Gamma \propto t^{-2}$.}
\label{GB-B}
\end{figure}

In our model, there are seven parameters.
As a benchmark case, hereafter we adopt $E_0=10^{52}$ erg, $\Gamma_0=100$,
$n_{\rm ISM}=1~\mbox{cm}^{-3}$, $p=2.2$, $\epsilon_{\rm e}=0.1$,
$\epsilon_B=0.1$, and $\eta=1$.
Figure \ref{GB-B} shows the evolution of $\Gamma \beta$
and $B'$ numerically obtained for the benchmark parameter set.
The deceleration starts at a slightly smaller radius than
the deceleration radius expressed by Equation (\ref{rdec}).
The evolution of $\Gamma \beta$ agrees with the adiabatic approximation
from Blandford--Mckee ($\Gamma \propto t^{-3/2}$ in the relativistic regime)
to Sedov--Taylor ($\beta \propto t^{-3/5}$ in the non-relativistic regime) phases.
The decay of the magnetic field also follows the evolution of $\Gamma \beta$
as expressed in Equation (\ref{B-field}),
though a slight deviation from the approximation $B' \propto \Gamma$
is seen below $\Gamma \beta<10$,
where the approximation $n'=4 \Gamma n_{\rm ISM}$ is not so accurate.

We also test the radiative case with $\epsilon_{\rm e}=0.9$
and $\epsilon_B=0.05$, where the shock-dissipated energy
is efficiently released by radiation.
The other parameters are the same as those in the benchmark case.
As shown by the dashed line in Figure \ref{GB-B},
the numerical result shows $\Gamma \propto t^{-2}$
in the relativistic regime, while the well-known formula
for the radiative shock \citep{bla76,sar98}
is $\Gamma \propto t^{-3}$.
The analytic formula in the radiative case is based on the approximation
$E_{\rm iso} \simeq \Gamma M_0 c^2$ neglecting the increase
of the inertia for $R<[3 M_0/(4 \pi m_{\rm p} n_{\rm ISM})]^{1/3}$
($\simeq 2.5 \times 10^{17}$ cm in this case).
Even for $\epsilon_{\rm e}=0.9$, however, the shocked ISM of mass $\Delta M$ adds
inertia $\geq (1-\epsilon_{\rm e})\Gamma \Delta M
=0.1 \Gamma \Delta M$, which is larger than $\Delta M$
for $\Gamma>10$.
Actually, for $t>3 \times 10^{6}$ s, the increase of the inertia cannot be negligible.
In addition, the faster decay of the magnetic fields leads to
the suppression of the radiative efficiency.
The increase of the inertia and decrease of the radiative efficiency
lead to $\Gamma \propto t^{-2}$
rather than $\Gamma \propto t^{-3}$ in this parameter set.

\begin{figure}[!ht]
\centering
\epsscale{1.0}
\plotone{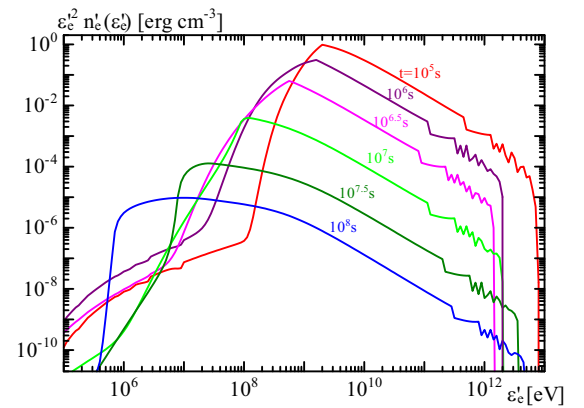}
\caption{Evolution of the electron energy distribution
in the shell frame for the benchmark case. The time $t$ labeled
in the figure is measured in the central engine frame.}
\label{E-evol}
\end{figure}

Figure \ref{E-evol} shows the evolution of $n'_{\rm e}(\varepsilon'_{\rm e})
\equiv N'_{\rm e}(\varepsilon'_{\rm e})/V'$
for the benchmark case.
In our numerical code, electrons are injected intermittently.
In the highest-energy region, the interval of the electron injection
is longer than the cooling time scale,
which results in the fluctuation of the electron distribution
as seen in the figure.
However, this does not practically affect the photon spectrum
because of the longer photon escape time than the electron injection interval.
The low-energy component below $10^8$ eV seen in the early stage
is down scattered particles with photons.

Initially the system is in the fast cooling regime.
The steady analytical solution for the fast cooling
is $n'_{\rm e}(\varepsilon'_{\rm e}) \propto \varepsilon'^{-2}_{\rm e}$.
In our time-dependent treatment, the injection rate increases with time,
so that the electron distribution below $\varepsilon'_{\rm min}$
(e.g. $5.6 \times 10^8$ eV at $t=10^{6.5}$ s)
is harder than the steady solution.
Equation (\ref{teq}) indicates that the electron distribution
should be expressed with the slow cooling approximation
for $t > 10^{6.5}$ s.
Actually, a sharp low-energy cutoff appears below $\varepsilon'_{\rm min}$
for $t \geq 10^{7}$ s.
The cooling break in the electron spectrum in the slow cooling regime
(see, e.g. $\sim 10^9$ eV at $t=10^{7.5}$ s)
is not so sharp;
the simple broken-power-law function is not appropriate for our results.

\begin{figure}[!ht]
\centering
\epsscale{1.0}
\plotone{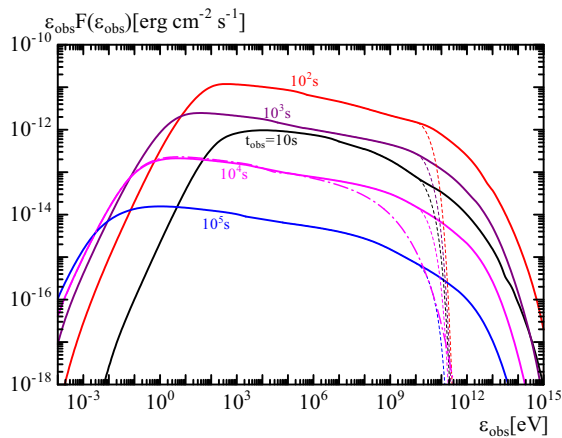}
\caption{Evolution of the photon spectrum for the benchmark case.
The source redshift is assumed as $z=2$.
The dashed lines are spectra including the effect of the intergalactic
$\gamma \gamma$-absorption.
The dot-dashed curve for $t_{\rm obs}=10^4$ s is the spectrum
obtained switching off the SSC emission artificially.}
\label{Ph-evol}
\end{figure}

\begin{figure}[!htb]
\centering
\epsscale{1.0}
\plotone{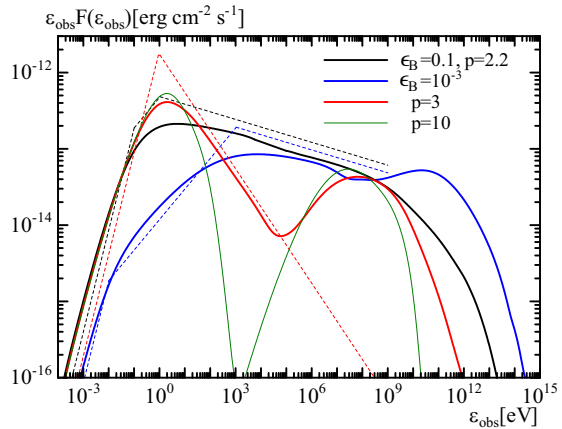}
\caption{Photon spectra at $t_{\rm obs}=10^4$ s with $z=2$
neglecting the intergalactic $\gamma \gamma$-absorption.
The black curve shows the benchmark case,
while the blue and red curves are results with the parameter sets
changing only one parameter from the benchmark parameter set:
$\epsilon_B=10^{-3}$ for the blue curve, and $p=3$ for the red curve.
The thin dashed lines are the synchrotron components
obtained analytically.
The thin green curve denotes the test calculation
with $p=10$ to mimic the monoenergetic injection,
where the other parameters are the same as the benchmark parameter set.}
\label{Ph-comp}
\end{figure}

The obtained photon spectrum for an observer in the benchmark case is rather simple
as shown in Figure \ref{Ph-evol}.
According to Equation (\ref{teq-obs}),
the photon spectrum must be the shape described with the slow cooling approximation
for $t_{\rm obs}>10^3$ s.
However, the spectra are smoothly curved around the peak,
so that it is hard to identify the spectral break at $\varepsilon_{\rm m}$
and $\varepsilon_{\rm c}$
(see Figure \ref{Ph-comp}).
In this parameter set, the synchrotron and SSC components
almost merge; the spectrum seems to consist of a single component.
As shown by the comparison of the solid and dot-dashed curves in Figure \ref{Ph-evol},
the SSC component dominates above 0.1 GeV for $t_{\rm obs}=10^4$ s.

At $t_{\rm obs}=10^4$ s, we compare the photon spectrum of the benchmark case
with the results for other parameter sets in Figure \ref{Ph-comp}.
When a parameter $\epsilon_B$ is reduced from 0.1 to $10^{-3}$
(see blue curve in Figure \ref{Ph-comp}),
$\varepsilon_{\rm c}$ grows into the X-ray range,
and the SSC component is clearly seen at the TeV energy range.
The analytical estimate implies $\gamma_{\rm c}/\gamma_{\rm m} \sim 300$
at this time.
Following \citet{sar01},
the ratio of the IC peak flux to the synchrotron one
is estimated as $(\gamma_{\rm c}/\gamma_{\rm m})^{(2-p)/2} \sqrt{\epsilon_{\rm e}/\epsilon_B}
\sim 6$,
while the numerical result shows a slightly dimmer IC flux than the synchrotron flux.
This discrepancy may be partially due to the Klein--Nishina effect,
but the time-dependent treatment apparently affects the flux ratio.
For the result of $p=3$ (see red curve in Figure \ref{Ph-comp}),
the soft synchrotron component makes the SSC component easier
to distinguish even for $\epsilon_B=0.1$ (see also the thin green line,
which corresponds to the monoenergetic injection).
Adopting Equations (\ref{em}), (\ref{ec}), and (\ref{Fmax}),
we also plot the analytic spectra of the synchrotron component in Figure \ref{Ph-comp},
where the maximum photon energy is simply assumed as $0.1 \Gamma/(1+z)$ GeV.
The numerical results
show curved spectra rather than the broken-power-law.
Those curved features are similar to the time-dependent calculations
in \citet[][see also \citet{pen14,uhm14}]{pet09}.
The analytic broken-power-law formula roughly reproduces
the overall spectral shape.
As shown in Figure \ref{Ph-comp}, however,
the analytic fluxes are slightly overestimated for $\varepsilon_{\rm obs}
\gtrsim \varepsilon_{\rm c}$
compared to our results.

\begin{figure}[!htb]
\centering
\epsscale{1.0}
\plotone{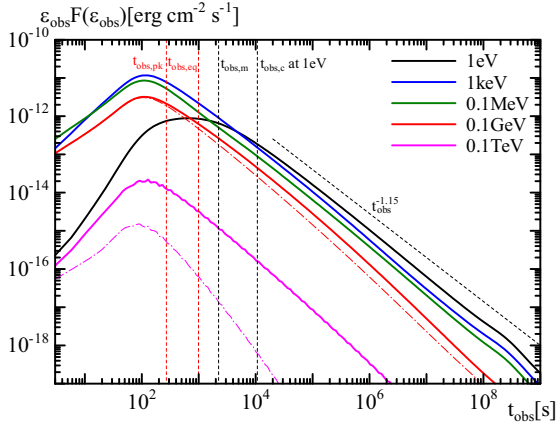}
\caption{Model light curves for the benchmark case with $z=2$.
The black, blue, green, red, and magenta curves are for 1 eV,
1 keV, 0.1 MeV, 0.1 GeV, and 0.1 TeV, respectively.
The dot-dashed curves are results
obtained by switching off the SSC emission artificially.
The analytically obtained typical times, $t_{\rm obs,pk}$ and $t_{\rm obs,eq}$,
are denoted by the vertical red dashed curves, while
$t_{\rm obs,m}$ and $t_{\rm obs,c}$ at 1 eV are denoted by
the vertical black dashed curves.
The analytic guiding line of $F \propto t_{\rm obs}^{-1.15}$
for the late stage is also plotted.
The flux at 0.1 TeV is suppressed by the intergalactic $\gamma \gamma$-absorption.}
\label{LC}
\end{figure}

The light curves for various photon energies are plotted in Figure \ref{LC}.
The X-ray peak time is 2.4 times earlier than Equation (\ref{tpk}).
If we adopt Equation (\ref{g0}) with the numerically obtained peak time
to estimate $\Gamma_0$,
this discrepancy leads to about 40\% larger $\Gamma_0$
compared to the actual value.
While Equation (\ref{g0}) is the same as the standard formula \citep{sar99,zha03,mol07},
the formula in \citet{lia10} is two times larger.
If our $t_{\rm obs,pk}$--$\Gamma_0$ relation obtained numerically is adopted,
the resultant $\Gamma_0$ becomes 2.8 times smaller than
the results in \citet{lia10}, in which a relation
$\Gamma_0 \simeq 182 (E_{\gamma,\mbox{iso}}/10^{52}\mbox{erg})^{0.25}$
is obtained from the afterglow onset times of 17 GRB samples.

Before the peak time, the X-ray flux grows as $F \propto t_{\rm obs}^{1.4}$,
while the simple analytical estimate 
leads to $F \propto t_{\rm obs}^2$ (see Equation (\ref{eqeLC})).
The fitting of the early X-ray light curves for 11 GRB samples by \citet{lia10}
shows a large scatter in the rising indices from 0.5 to $\sim 10$.
At $\varepsilon_{\rm obs}=1$ eV, which is below $\varepsilon_{\rm c}$
in the early stage, the rising index in our calculation is about 2.2,
which is also smaller than the analytical estimate $11/3$ (Equation (\ref{eqeLC2})).

The standard analytic model \citep{sar98,sar99} predicts the evolution
for the 1eV light curve as
$F \propto t^{1/6}$ for
$t_{\rm obs,pk} < t_{\rm obs} < t_{\rm obs,eq}$,
$F \propto t^{1/2}$ for
$t_{\rm obs,eq} < t_{\rm obs} < t_{\rm obs,m}$,
and $F \propto t^{3(1-p)/4}=t^{-0.9}$ for
$t_{\rm obs,m} < t_{\rm obs} < t_{\rm obs,c}$
(see Equations (\ref{eqsfast}) and (\ref{Fl-ana})).
Since the photon spectrum is curved around $\varepsilon_{\rm m}$
or $\varepsilon_{\rm c}$, our result does not show
such sharp breaks in the 1eV light curve.
For $t>t_{\rm obs,c}$, the light curves below MeV
seem consistent with the decay $F \propto t^{-1.15}$
depicted in Equation (\ref{Fl-ana}).

We have optimistically assumed the Bohm limit $\xi=1$
to maximize the maximum electron energy $\varepsilon'_{\rm max}$.
As a result, below $0.1 \Gamma (1+z)^{-1}$ GeV, the synchrotron emission dominates.
Thus, even if we neglect the SSC emission, the light curves
below 0.1 GeV are almost unchanged,
while the 0.1 TeV emission is greatly suppressed.

\begin{figure}[!ht]
\centering
\epsscale{1.0}
\plotone{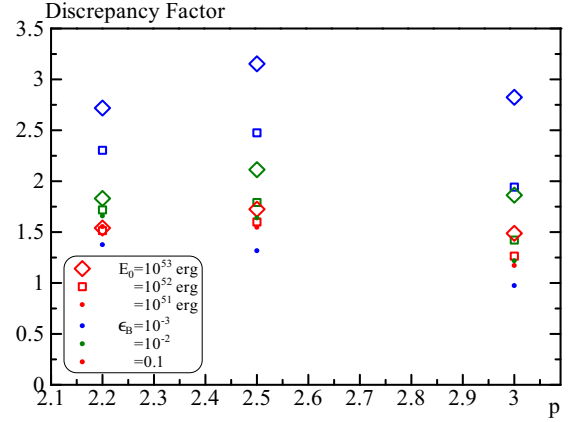}
\caption{Discrepancy factors between the analytical and numerically calculated fluxes
at 1 keV and $t_{\rm obs}=10^4$ s for 27 parameter sets
with $z=2$. Since the magnetic field is strong enough as $\epsilon_B \geq 10^{-3}$,
1 keV is above $\varepsilon_{\rm c}$ ($>\varepsilon_{\rm m}$) in most of the cases.}
\label{disc}
\end{figure}

As we have mentioned, the analytic formula reviewed in section \ref{sec:ana}
tends to overestimate the X-ray flux (see Figure \ref{Ph-comp}).
Here, we define the ``discrepancy factor'' of the analytical flux
based on Equations (\ref{em}), (\ref{ec}), and (\ref{Fmax}) as the ratio of
\begin{eqnarray}
\frac{\mbox{Analytically Estimated Flux}}{\mbox{Numerically Obtained Flux}},
\label{df}
\end{eqnarray}
at 1 keV at $t_{\rm obs}=10^4$ s assuming $z=2$.
We change the three model parameters as $E_{52}=0.1$, 1, and 10,
$\epsilon_B=0.1$, $10^{-2}$, and $10^{-3}$,
and $p=2.2$, $2.5$, and $3.0$.
In total we test $3 \times 3 \times 3=27$ models,
keeping the other parameters as $\Gamma=100$,
$n_{\rm ISM}=1~\mbox{cm}^{-3}$, $\epsilon_{\rm e}=0.1$,
and $\eta=1$.
The discrepancy factors are shown in Figure \ref{disc}.
Although the normalization of the electron number $\dot{N'_0}$
is proportional to $(p-1)$, which is neglected in the estimate of $F_{\rm max}$
in Equation (\ref{Fmax}),
our results do not show a clear dependence on $p$
in the discrepancy factors.
In the cases where the SSC emission is efficient,
namely, smaller $\epsilon_B$ and larger $E_0$,
the discrepancy factor tends to be large.
This tendency agrees with the analytical discussion
in \citet{ben16}.

\begin{figure}[!ht]
\centering
\epsscale{1.0}
\plotone{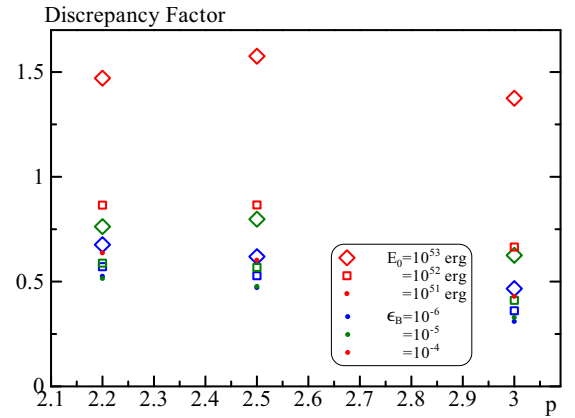}
\caption{Same as Fig. \ref{disc} but for smaller values of the magnetic parameter
as $\epsilon_B=10^{-4}$--$10^{-6}$.
In most of those cases, 1 keV is in the range between $\varepsilon_{\rm m}$
and $\varepsilon_{\rm c}$.}
\label{disc2}
\end{figure}

Note that the normalization of the flux in Equation (\ref{Fmax})
is basically the same
as used in \citet{llo04} to estimate the kinetic energy $E_0$
in the afterglow phase.
\citet{llo04} raised the problem that
the prompt gamma-ray emission is too efficient
compared to the remnant kinetic energy at the afterglow onset.
The analytical formulation reviewed in section \ref{sec:ana} shows that
the synchrotron flux
in the highest-energy region is proportional
to $F_{\rm max} \varepsilon_{\rm c}^{1/2} \varepsilon_{\rm m}^{(p-1)/2}
\propto E_0^{(p+2)/4}$ ($E^{1.05}$ for $p=2.2$).
Therefore, $E_0$ obtained from the analytic formula
results in an underestimate of $E_0$ by a factor
close to the discrepancy factors shown in Figure \ref{disc}.

\begin{figure}[!htb]
\centering
\epsscale{1.0}
\plotone{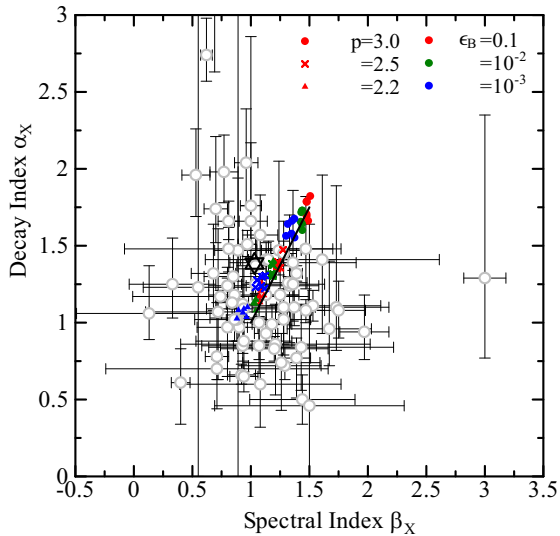}
\caption{Scatter plot of the spectral index vs. decay index
for 54 models with higher values of $\epsilon_B$ ($\geq 10^{-3}$).
To avoid complication,
we do not distinguish the symbols for the differences in $E_0$
and $\Gamma_0$. The solid line is the analytic closure relation
in the fast cooling regime ($\varepsilon_{\rm c}<\varepsilon_{\rm obs}$)
for the adiabatic case
between $p=2$ and 3.
The star symbol is the radiative case
($\epsilon_{\rm e}=0.9$ and $\epsilon_B=0.05$;
the other parameters are the same as those in the benchmark case).
The observed samples (gray) with error bars
are taken from \citet{wil07}.}
\label{Cls}
\end{figure}

Here we have assumed $\epsilon_B \geq 10^{-3}$ so that
the X-ray-emitting electrons are in the fast cooling regime
at $t_{\rm obs}=10^{4}$ s.
Namely, the X-ray band is above both $\varepsilon_{\rm m}$
and $\varepsilon_{\rm c}$.
However, the recent broadband
observations \citep[e.g.][]{kum09,lem13,san14,zha15}
suggest much lower magnetization.
\citet{ben15,ben16} pointed out that the high-efficiency problem in the prompt emission
will be resolved by a very small $\epsilon_B$.
In this case, the X-ray-emitting electrons are in the slow cooling regime,
and the synchrotron flux in the X-ray band can be suppressed by
the IC cooling.
Those two effects may lead to a wrong estimate of $E_0$, if we adopt
the usual fast cooling formula for the X-ray-emitting electrons.

We also test the discrepancy factor for $\epsilon_B=10^{-4}$--$10^{-6}$
as shown in Figure \ref{disc2}.
In most of those cases, the X-ray band is between $\varepsilon_{\rm m}$
and $\varepsilon_{\rm c}$.
The analytical broken-power-law formula tends to underestimates the flux
for $\varepsilon_{\rm m}<\varepsilon_{\rm obs}<\varepsilon_{\rm c}$
contrary to the fast cooling regime ($\varepsilon_{\rm c}<\varepsilon_{\rm obs}$).
Figure \ref{disc2} shows that the analytical formula underestimate the flux
by a factor of 1.2--3.

This implies that the total energy obtained from the analytical formula
tends to be overestimated for smaller $\epsilon_B$, which will worsen
the high-efficiency problem in the prompt emission.
However, if we misunderstand the X-ray energy range as the fast cooling regime,
the estimate of $E_0$ can be less than $1$\% of the actual energy
depending on the parameters.
This misinterpretation can be a major factor that leads to an underestimate of $E_0$
as pointed out by \citet{ben15,ben16}.
In such cases, the discrepancy shown in Figure \ref{disc2} seems
negligible.

As is understood from Figures \ref{disc} and \ref{disc2},
the suppression of the X-ray flux by the IC cooling becomes maximum at $\epsilon_B \sim 10^{-3}$.
For an extremely small $\epsilon_B$, though the IC cooling becomes relatively dominant
compared to the synchrotron cooling, the cooling effect itself is negligible.
Therefore, the synchrotron emissivity is not largely affected by the radiative cooling.

We also test the X-ray closure relation between the decay index $\alpha_{\rm X}$
($F \propto t_{\rm obs}^{-\alpha_{\rm X}}$)
and spectral index $\beta_{\rm X}$ ($F \propto \varepsilon_{\rm obs}^{-\beta_{\rm X}}$).
The analytical formula of Equation (\ref{Fl-ana}) indicates
$\alpha_{\rm X}=(3p-2)/4$ and $\beta_{\rm X}=p/2$ for $\varepsilon_{\rm obs}>
\varepsilon_{\rm c}$.
This implies the closure relation $(3/2)\beta_{\rm X}-\alpha_{\rm X}=1/2$.
We may expect deviation from the closure relation in the numerical results.
In addition to the 27 models in Figure \ref{disc},
we change the initial Lorentz factor as $\Gamma=100$ and 300
and obtain $\alpha_{\rm X}$ and $\beta_{\rm X}$ at 1 keV and $t_{\rm obs}=10^4$ s
assuming $z=2$.
The numerical results for all 54 models are plotted in Figure \ref{Cls}.
In most cases, the results slightly deviate from the analytical relation
for the corresponding $p$.
As $\epsilon_B$ decreases, the spectrum tends to be hard,
and the flux decay tends to be shallow.
Nevertheless, our numerical results in Figure \ref{Cls} distribute
along the analytic closure relation.
Those points are slightly above the closure relation systematically.

The result for the radiative model ($\epsilon_{\rm e}=0.9$ and $\epsilon_B=0.05$)
is also plotted in Figure \ref{Cls}.
In the analytic model \citep{sar98},
$\alpha_{\rm X}=(6p-2)/7=1.6$ and $\beta_{\rm X}=1.1$ in this case.
The obtained values are $(\alpha_{\rm X},\beta_{\rm X})=(1.38,1.03)$.
The shallower decay of $\Gamma$ than the analytic model as shown in Figure \ref{GB-B}
causes the shallower flux decay.
Even in this extreme model, $\alpha_{\rm X} \sim 2$ is hard to realize.

\begin{figure}[!htb]
\centering
\epsscale{1.0}
\plotone{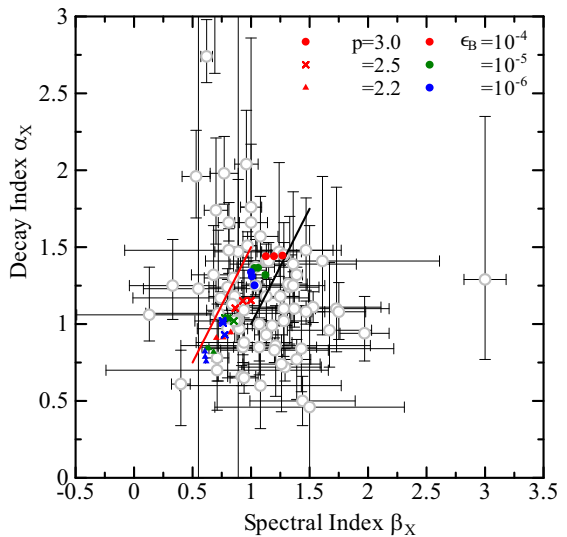}
\caption{Same as Fig. \ref{Cls} but for
27 models with $\epsilon_B \leq 10^{-4}$ and $\Gamma=100$.
The red solid line is the analytic closure relation
in the slow cooling regime
($\varepsilon_{\rm m}<\varepsilon_{\rm obs}<\varepsilon_{\rm c}$)
between $p=2$ and 3.}
\label{Cls2}
\end{figure}

As we have mentioned, for $\epsilon_B \leq 10^{-4}$,
the X-ray band is in the slow cooling regime in most cases.
In this case, the closure relation becomes
$(3/2)\beta_{\rm X}-\alpha_{\rm X}=0$.
As shown in Figure \ref{Cls2}, the numerical results show slight deviation
from the closure relation (the red line),
while the distribution of the decay indices $\alpha_{\rm X}$
is consistent with the analytical formula.
The $\alpha_{\rm X}$--$\beta_{\rm X}$ distribution in this case
does not contradict the observed distribution.

However, the large scatter in the samples in \citet{wil07}
is not explained by the model with constant $n_{\rm ISM}$,
$\epsilon_{\rm e}$, $\epsilon_B$, or $\eta$.
Our model does not take into account the shallow decay phase,
which may require the energy injection with a longer timescale than $t_{\rm obs,pk}$.
Though some additional parameters with respect to the energy injection
may resolve the problem,
such a complex model is beyond the scope in this paper.

\section{Application to GRB 130427A}
\label{sec:130}

\begin{figure*}[!htb]
	\begin{center}
		$\begin{array}{cc}
		\includegraphics[width=0.4\textwidth]{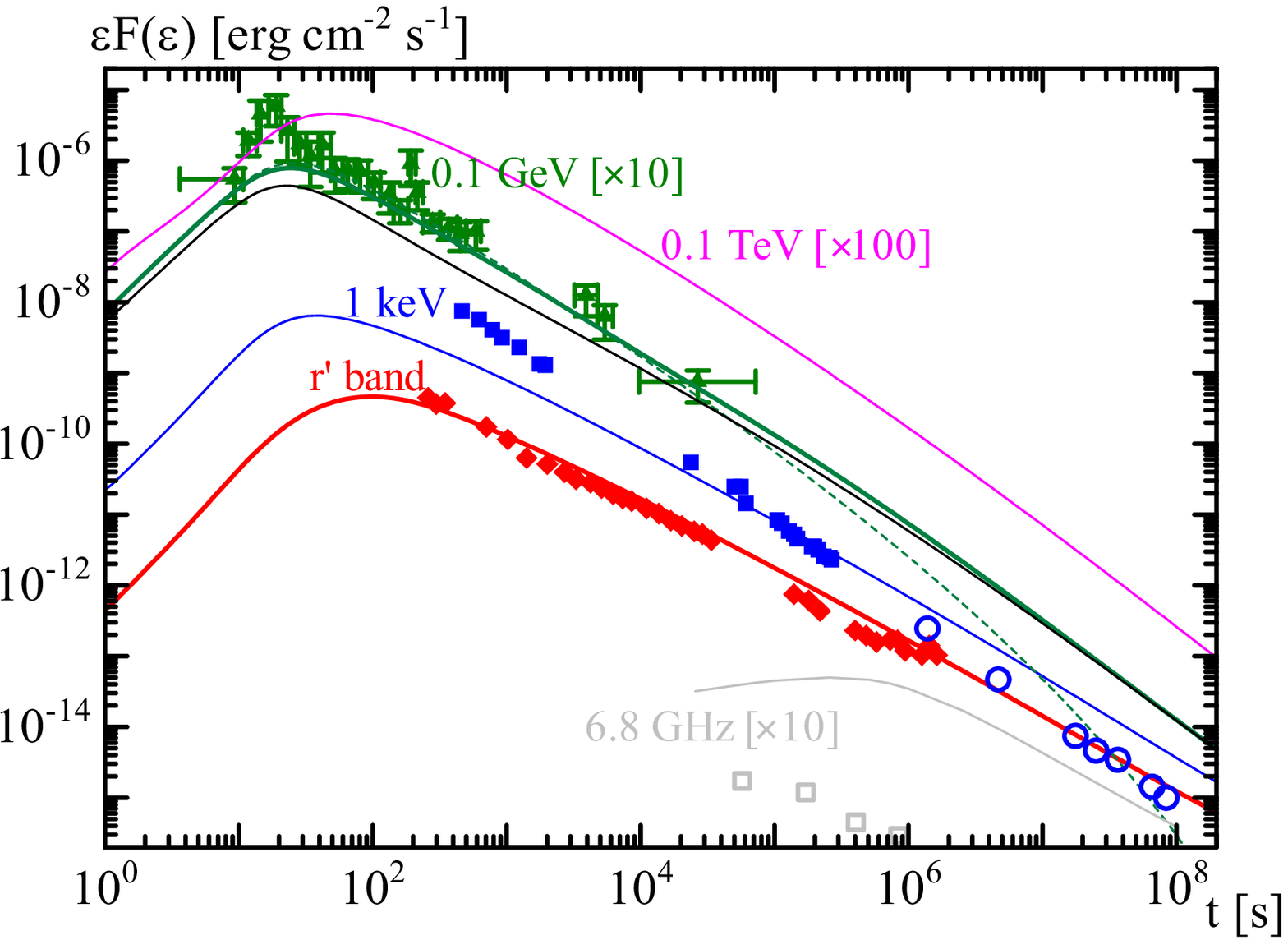} &
		\includegraphics[width=0.4\textwidth]{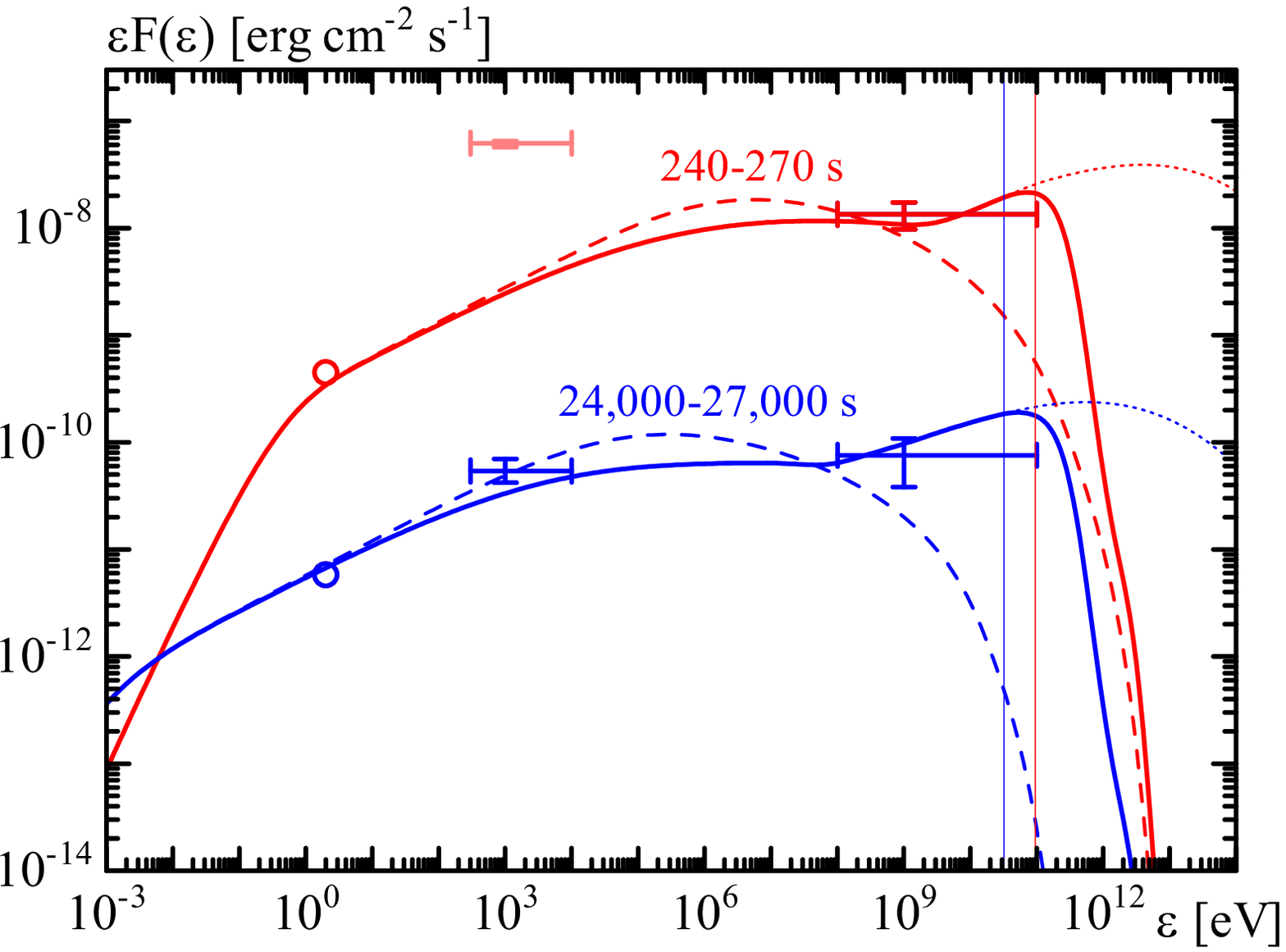}
		\end{array}$
	\end{center}
	\caption{Light curves (left) and spectra (right) of model A for GRB 130427A.
The data points are taken from \citet{ack14} and \citet{mas14}.
The late X-ray data (open circles) are taken from \citet{dep16}.
Left: to plot the fluxes at 0.1 GeV and keV,
the photon index is assumed to be 2.0.
The model light curves are plotted with solid purple (0.1 TeV), green (0.1 GeV),
blue (keV), red ($r'$ band), and gray (6.8 GHz) lines.
The thin black line represents a 0.1 GeV light curve
with the same parameters as for model A but with $\xi=100$.
The green dashed line is also the 0.1 GeV light curve of model A,
but switching off the SSC emission artificially.
Right: the data points for 0.1--100 GeV flux
are averaged values over 237--318 s and 9,720--72,200 s, respectively.
The thick dashed lines show spectra obtained
by switching off the SSC emission artificially.
The thin dotted lines are spectra neglecting the effect of the intergalactic
$\gamma \gamma$-absorption.
The red and blue vertical lines indicate the photon energy detected
at the observation times of $244$ s and 34,400 s, respectively.
}
	\label{fig:modelA}
\end{figure*}

GRB 130427A \citep{ack14,mas14} is a very nearby GRB ($z=0.34$)
with a very large isotropic energy release \citep[$8.5 \times 10^{53}$ erg;][]{per14}
in gamma-ray.
Surprisingly, the X-ray afterglow flux
is well fitted by a simple power-law of $t^{-1.309}$
(hereafter we omit the subscript ``obs'') as far as $8 \times 10^7$ s
without a signature of the jet break \citep{dep16}.
Thanks to the very large fluence, a long-lasting GeV emission
as far as $7 \times 10^4$ s was detected with {\it Fermi}.
The most enigmatic problem in this GRB is the detection of a 32 GeV photon
at $t \simeq 3 \times 10^4$ s.
The maximum synchrotron photon energy is limited by
the balance between the energy loss and gain
as $\sim 0.1$ GeV irrespective of the magnetic field.
While the relativistic motion can boost the photon energy in the early stage,
the Lorentz factor should be significantly reduced
at the arrival time of the 32 GeV photon.
Therefore, the 32 GeV photon may be emitted via SSC process
\citep{fan13,liu13}.
However, the GeV light curve is well fitted by a single power-law
in the late phase and does not show the signature
of the transition from synchrotron to SSC.

\begin{table}[!hbtp]
	\caption{Model parameters for the afterglow emission of GRB 130427A.}
	\begin{center}
		\begin{tabular}{lrrrrrrr}
			\hline\hline
			 & $E_0$ (erg) & $\Gamma_0$ & $\epsilon_{\rm e}$ & $\epsilon_B$ & $n_0$ & $p$ &
            DF\footnote{Discrepancy factor defined by Eq. (\ref{df}).} \\
			\hline
			Model A & $2.0 \times 10^{55}$ & $350$ & $0.03$ & $10^{-6}$ & $1.0$ & $2.35$ & 0.66\\
			Model B & $4.5 \times 10^{55}$ & $400$ & $0.015$ & $10^{-5}$ & $1.0$ & $2.60$ & 1.6\\
			Model C & $4.5 \times 10^{54}$ & $400$ & $0.23$ & $10^{-5}$ & $0.1$ & $2.60$ & 2.7 \\
			\hline\hline
		\end{tabular}
	\end{center}
    \tablecomments{The parameter $\eta$ is fixed as unity. The Bohm factor $\xi$,
which scales the acceleration time, is also assumed as unity.}
	\label{param_tab}
\end{table}

The decay indices of the optical and X-ray light curves are different,
which implies that there are difficulties in
the standard ISM model with a single emission component.
Especially for the early stage of the afterglow,
complicated models including a reverse shock component
with a stellar-wind profile \citep{las13,per14,ves14},
spine-sheath-like two-component jets \citep{vHor14},
temporally evolving microscopic parameters \citep[$\eta$,
$\epsilon_{\rm e}$ and $\epsilon_B$;][]{mas14},
or a non-standard radial density profile \citep{kou13,vHor14}
have been attempted.
Nevertheless, we adopt our single-component model focusing on
the GeV--TeV emission in the early phase;
our numerical method is ideal to calculate the GeV--TeV light curve
with SSC emission consistently
as explained in the previous section.
We have tested three models, whose parameters are summarized in Table \ref{param_tab}.

As we have mentioned, it may be difficult
to reproduce both the X-ray and optical light curves by a single emission component
with constant microscopic parameters.
In model A, we give weight to the optical light curve as shown in Figure
\ref{fig:modelA}.
For $t \lesssim 10^4$ s, the model flux in X-ray is dimmer than the observed one.
Another emission component such as the reverse shock may be required
to agree with the early X-ray light curve.
The small value of the parameter $\epsilon_{\rm e}$ leads
to a large value of $E_0$.
If we adopt a higher $\epsilon_{\rm e}$, the high initial $\gamma_{\rm m}$
makes the peak time of the optical light curve
delayed compared to the observed onset time (see Model C).
As the spectrum for 240--270 s in Figure \ref{fig:modelA} shows,
to generate such bright synchrotron flux as the early X-ray data indicate,
$\varepsilon_{\rm m}$ should be in the X-ray energy range.
However, such a high $\varepsilon_{\rm m}$ contradicts the decaying
flux of the optical emission at this stage.

The 0.1 GeV light curve is well reproduced by our model.
The model curve shows a smooth-power-law-like behavior.
Around 0.1 GeV, both the synchrotron and SSC emissions contribute.
Even if we artificially turn off the SSC emission (see the green dashed line
in the left panel of Figure \ref{fig:modelA}),
the 0.1 GeV emission due to synchrotron yields a single power-law
light curve until $10^5$ s.
Therefore, the 0.1 GeV range is not so ideal
to find the switching from synchrotron to SSC in the light curve.

\begin{figure*}[!htb]
	\begin{center}
		$\begin{array}{cc}
		\includegraphics[width=0.4\textwidth]{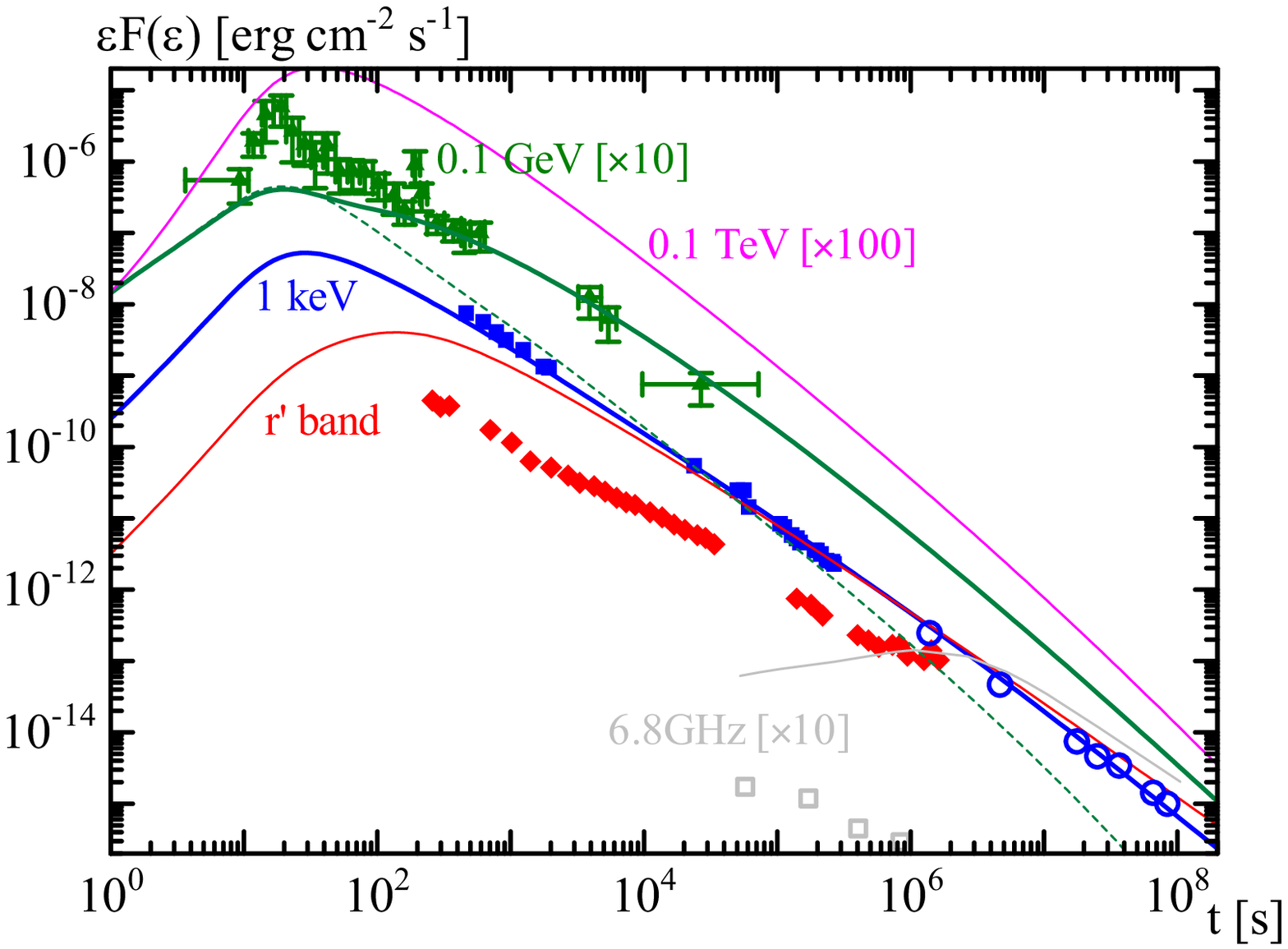} &
		\includegraphics[width=0.4\textwidth]{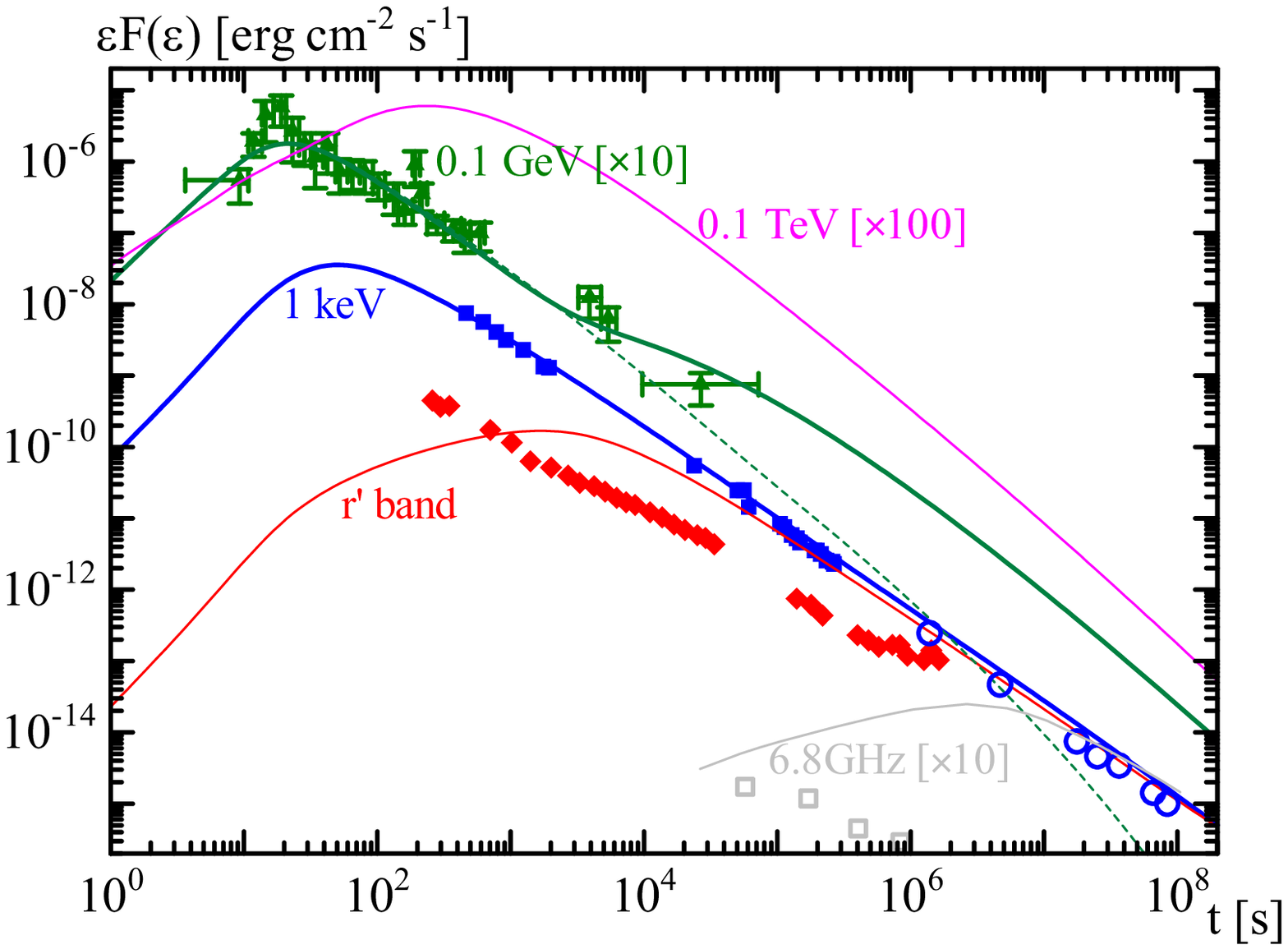}
		\end{array}$
	\end{center}
	\caption{Light curves of model B (left) and model C (right) for GRB 130427A.
The data points are the same as in Fig. \ref{fig:modelA}.
The model light curves are plotted with solid purple (0.1 TeV), green (0.1 GeV),
blue (keV), and red ($r'$ band) lines.
The green dashed lines are 0.1 GeV light curves,
but switching off the SSC emission artificially.
}
	\label{fig:modelBC}
\end{figure*}

The detections of the 95 and 32 GeV photons
are not explained by synchrotron emission, as shown in the right panel
of Figure \ref{fig:modelA} (see dashed lines for the model without SSC,
and red and blue vertical lines for the energies of the detected photons).
Even in the early period of $t \sim 200$--$300$ s,
photons above 10 GeV are emitted via SSC in model A.
The predicted 0.1 TeV light curve (purple) is also smooth
and lasts a long time.
Even at $t=10^4$ s, the flux at 0.1 TeV is about
$5 \times 10^{-10}~\mbox{erg}~\mbox{cm}^{-2}~\mbox{s}^{-1}$,
which can be detected with CTA with a time resolution of a few hundred seconds
\citep{fun13,ino13}.

While we have adopted the Bohm factor as unity in Model A,
the thin black line in the right panel of Figure \ref{fig:modelA} shows
the 0.1 GeV light curve with $\xi=100$.
In this conservative model, the dominant emission process in the 0.1 GeV range
is replaced from synchrotron to SSC in the later phase.
However, the 0.1 GeV light curve is smooth even in this case.

Next, giving weight to the X-ray light curve
rather than the optical one, let us try to find an acceptable model.
The steeper slope of the X-ray light curve leads to a larger $p$,
which makes the spectrum softer.
Model B is an example of our results that agree with
the observed X-ray and 0.1 GeV light curves (see left panel of Figure \ref{fig:modelBC}).
As the green dashed line indicates, the emission at 0.1 GeV for $>10^2$ s is dominated
by SSC in model B.
The soft spectrum results in brighter optical flux than observed.
\citet{mas14} claimed that the optical extinction is negligible
from the SED analysis.
The relatively dim optical fluxes and the steep X-ray decay
seem difficult to explain simultaneously by a single source model
with constant microscopic parameters.
However, interestingly, this model omitting the effect of the jet break is consistent with
the X-ray light curve as far as $\sim 10^8$ s,
though no signature of the jet break challenges the standard afterglow model
and implies a very large energy release for this GRB.
As \citet{dep16} pointed out, the previous complex models
\citep{kou13,las13,pan13,mas14,per14,vHor14} have difficulties reconciling
the observed long-lasting X-ray emission.
Although we focus on the early afterglow rather than the late one,
the physical parameters may be close to
those of model B in the late phase.
However, the predicted radio flux at 6.8 GHz is significantly brighter than
the observed flux \citep{mas14}.

Another example is model C, whose light curves are shown
in the right panel of Figure \ref{fig:modelBC}.
By increasing $\epsilon_{\rm e}$ and decreasing $n_0$,
the required energy $E_0$ is drastically suppressed
compared to models A and B.
In model C, $\varepsilon_{\rm m}$ is kept higher than the optical range
for a long time.
The resultant optical light curve shows a late peak time,
which seems inconsistent with the simple power-law decay
of the observed light curve.
The high $\gamma_{\rm m}$ in this model leads to the dominant
contribution of the synchrotron emission at 0.1 GeV as far as
a few times $10^3$ s, after which the contribution of SSC emission modulates
the 0.1 GeV light curve.
This deviation from single power-law in the 0.1 GeV light
is still within the observational errors.

All the models in Table \ref{param_tab} have a very small value of $\epsilon_B$,
which agrees with the results of recent studies \citep{kum09,lem13,san14,ben15,zha15}.
In spite of the small $\epsilon_B$, the discrepancy factors in the X-ray flux
are of the order of unity (see Table \ref{param_tab}).
The initial magnetic fields of models A--C are 0.14, 0.49, and 0.16 G, respectively.
A shock-compressed CSM magnetic field is
only $4 \Gamma_0 B_{\rm CSM}\simeq 1.4 (\Gamma_0/350) (B_{\rm CSM}/\mu\mbox{G})$ mG.
Even for those small $\epsilon_B$,
an amplification mechanism of the magnetic field
is required \citep[see, e.g.][]{barD14}.

\section{Conclusions}
\label{sec:conc}

In order to simulate the GRB afterglow emission,
we have calculated the temporal evolutions of the energy distributions
of electrons and photons in the shell relativistically propagating
in the ISM.
Physical processes such as the deceleration of the shell, photon escape,
adiabatic cooling, and transformations of observables into the observer frame
are consistently dealt with in our numerical code.
Given the initial Lorentz factor $\Gamma_0$, the onset time of the afterglow
in our results is significantly earlier than the previous analytical
estimate.
The uncertainty in the initial Lorentz factors
obtained from the onset time may be larger than previously thought.
When we mimic the radiative case by adopting an extreme value $\epsilon_{\rm e}=0.9$,
the results show $\Gamma \propto R^{-2}$ and $F \propto t_{\rm obs}^{-1.4}$,
which are significantly different from the conventional formulae.

In the fast cooling case, our results show that
the electron spectrum for $\gamma_{\rm c}<\gamma
< \gamma_{\rm m}$ is significantly curved and harder than the analytical estimate
owing to the evolution of the injection rate.
The spectral shape is highly curved around the typical energies $\varepsilon_{\rm m}$
and $\varepsilon_{\rm c}$.
While the peak flux at $\varepsilon \sim \varepsilon_{\rm m}$ or $\varepsilon_{\rm c}$
is lower than the analytical estimate
with the broken-power-law approximation,
the discrepancy of the X-ray flux with the analytical synchrotron formula
is not so large.
The total energy obtained by fitting the
observed light curves with the analytical formula
may be underestimated by a factor of three
or less.
However, as \citet{ben15,ben16} pointed out,
if we misunderstand that the X-ray-emitting electrons
are in the fast cooling regime despite $\epsilon_B \ll 10^{-3}$,
the total energy can be highly underestimated.
This may resolve the high-efficiency problem in the prompt emission
\citep{llo04}.

Our results show that even if the emission mechanism
is switching from synchrotron to SSC,
the gamma-ray light curves can be a smooth power-law,
especially for the electron index of $p \simeq 2$--2.5.
Note that we have not intentionally adjusted the parameters to suppress
the light-curve signature of the switching from synchrotron to SSC.
In most cases with fiducial parameter sets, 
it is difficult to find the time at which SSC starts contributing
from only light curves.

Given the electron spectral index $p$, the SSC contribution
makes the photon spectrum
slightly harder than the expectation from the synchrotron formula.
We have tested 54 models changing the parameters.
The numerically obtained spectral index and decay index
are scattered, but distribute along the analytical closure relation.
To explain GRBs whose indices largely deviate from the closure relation,
the evolution of the microscopic parameters
may be required.

With our method, we have fitted the light curves of GRB 130427A,
in which high-energy photons beyond the synchrotron limit
were detected.
Although our single-source model with constant microscopic parameters
does not reproduce all the observed behaviors in multiple wavelengths,
the combination of the synchrotron and
SSC emissions from the external shock can consistently explain
the smooth 0.1 GeV light curve and the detections of
95 and 32 GeV photons at $t=244$ s and 34,400 s, respectively.
As long as $\epsilon_B \ll 1$, as the recent studies suggested,
10--100 GeV SSC emission will be expected to be detected with CTA
\citep[see][for a conservative estimate of the detection rate]{vur16}.

\acknowledgements

First, we appreciate the valuable advise by the anonymous referee.
This work is supported by Grants-in-Aid for Scientific
Research nos. 15K05069, 16K05291 (K.A.), and 15K05080 (Y.F.) from the Ministry
of Education, Culture, Sports, Science and Technology
(MEXT) of Japan.

\end{document}